\documentstyle[preprint,aps,prl,epsf]{revtex}   
\begin{document}
\title{{\bf Thermal unfolding of proteins}}
\author{{\bf Marek Cieplak and Joanna I. Su{\l}kowska}}

\address{
Institute of Physics, Polish Academy of Sciences,
Al. Lotnik\'ow 32/46, 02-668 Warsaw, Poland}

\maketitle



\vspace*{1cm}

\begin{abstract}
{Thermal unfolding of proteins is compared to folding and mechanical 
stretching in a simple topology-based
dynamical model. We define the unfolding time and demonstrate
its low-temperature divergence. Below a characteristic temperature,
contacts break at separate time scales and unfolding
proceeds approximately in a way reverse to folding. Features
in these scenarios agree with experiments and atomic simulations
on titin.}
\end{abstract}

\newpage

Conformational changes in proteins occur in a variety of processes
such as folding, mechanically induced stretching, chemical denaturation,
and thermally induced unfolding. Standard molecular dynamics simulations
of these processes cover nanosecond intervals which usually  misses
the relevant time scales by at least six orders of magnitude.
One may gain insights into the  long time scale conformational
dynamics by considering effective coarse-grained models. Among these,
the simplest and yet often succesful are the
topology based models. They are constructed based on the knowledge of
the experimentally established native conformations \cite{Goabe,Stakada}.
The topology based models offer a possibility to study various 
processes within one unified approach, an opportunity to explore
relationships between them, and an easy way to determine dependence on
parameters, such as the temperature.\\

In this Letter, we focus on thermal unfolding. This phenomenon is often
invoked in theoretical searches for a molecular interpretation
of the transition state for the folding process \cite{Pande2}. 
The transition
state is probed experimentally through the protein-engineering-based
so called $\phi$-value analysis \cite{Fersht}. The assumptions 
underlying the theoretical search for the transition state by simulating  
thermal unfolding are that unfolding should proceed in a way that
is reverse to folding and that the transition state should be quickly
accessible from the native state, especially if high temperatures
are applied (even up to 200$^o$C which in itself may, however, 
alter the free energy landscape of a protein significantly).
Additional assumptions attempt to relate
the transition state to "large structural changes" \cite{Daggett,Daggett1}
in an unfolding evolution of a
protein -- a point recently assessed in Ref. \cite{ProteinScience}.
Here, we analyse thermal unfolding within the topology-based
model as implemented in Refs. \cite{biophysical} and 
\cite{thermtit}. We characterize unfolding
at various temperatures by determining unfolding times and 
by providing scenarios of unfolding.
We show that there is a characteristic temperature, $T_{\Omega}$,
associated with unfolding above which rupturing of bonds
occurs simultaneously (on an average)
at all sequential separations.
We show that the unfolding times diverge on lowering
the temperature and that below $T_{\Omega}$, 
the unfolding process runs in reverse to folding as monitored
at an optimal folding temperature. Some of our predictions regarding
scenarious of the conformational changes are found to be consistent with
experimental findings.\\

We consider several model proteins with a special emphasis on the
I27 globular domain of the muscle protein titin (the Protein Data
Bank \cite{PDB} code 1tit). Mechanical stretching of
this protein has been extensively studied in experiments involving
atomic force microscopy \cite{Gaub,Marszalek,Fowler2002} and there is also
some information about its folding \cite{Fowler2001,Geierhaas}. 
Furthermore, we have already studied it undergoing both processes through 
molecular dynamics simulations within the topology based 
model \cite{thermtit,Robbins,Pastore,calmodulin}.\\

This model can be outlined as follows. The protein is represented
by the C$^{\alpha}$ atoms that are tethered by harmonic potentials
with minima at 3.8 {\AA}. The native contacts are described by the
Lennard-Jones potentials
$V_{ij} =
4\epsilon \left[ \left( \frac{\sigma_{ij}}{r_{ij}}
\right)^{12}-\left(\frac{\sigma_{ij}}{r_{ij}}\right)^6\right]$, where
the length parameters $\sigma _{ij}$ are chosen so that the potential
minima correspond, pair-by-pair, to the native distances between
the C$^{\alpha}$ atoms $i$ and $j$. Which amino acids form native contacts
is determined through atomic overlaps as described by
Tsai et al. \cite{Tsai}.
The non-native contacts are described by repulsive cores of
$\sigma =5$ {\AA}. The energy parameter, $\epsilon$, is taken
to be uniform and its effective value appears to be of order
900 K, at least for titin. The optimal folding temperature,
$T_{min}$, for I27
has been found to correspond to the reduced temperature
$\tilde{T}=k_BT/\epsilon \;$ of 0.275 \cite{thermtit}
($k_B$ is the Boltzmann constant and $T$ is temperature)
which is close to the room temperature value of $\tilde{T}$=0.3.
In our stretching simulations, both ends of the protein are
attached to harmonic springs of elastic
constant $k$=0.12 $\epsilon /${\AA}$^2$ which is close to the
values corresponding to the elasticity of experimental cantilevers.
The free end of one of the two springs is constrained while the
free end of the second spring is pulled at constant speed, $v_p$,
along the initial end-to-end position vector. We focus on
$v_p$ of 0.005 {\AA}$/\tau$, where
$\tau =\sqrt{m \sigma^2 / \epsilon} \approx 3$ps
is the characteristic time for the
Lennard-Jones potentials. Here, $\sigma=5${\AA} is a typical value
of $\sigma_{ij}$ and $m$ is the average mass of the amino acids.
Thermostating is provided by the Langevin noise which also
mimics random kicks by the implicit solvent.
An equation of
motion for each C$^{\alpha}$ reads
$m\ddot{{\bf r}} = -\gamma \dot{{\bf r}} + F_c + \Gamma $, where
$F_c$ is the net force on an atom due to the molecular potentials.
The damping constant $\gamma$ is taken to be equal to $2m/\tau$
and the dispersion of the random forces is equal to
$\sqrt{2\gamma k_B T}$.
This choice of $\gamma$ corresponds to a situation in which the
inertial effects are negligible \cite{biophysical}
but the damping action is not yet as strong as in water.
Increasing $\gamma$ twentyfold results in a twentyfold increase in the
time scales bringing the typical value of $v_p$ within two orders of
magnitude of the experimental pulling speeds \cite{thermtit}
and correspondingly longer folding times \cite{biophysical}.
The equations
of motion are solved by a fifth order predictor-corrector scheme.\\

The top two panels of Figure 1 illustrate what happens to distances
between two pairs of amino acids that make native contacts when
submitting the I27 domain of titin to a very high reduced 
temperature of 1.1. The broken line corresponds to the distance of
1.5$\sigma _{ij}$ that is considered as a qualitative treshold for the
amino acids staying or not staying in a contact 
\cite{biophysical,Robbins}. It is seen that the treshold line
is being crossed repeatedly due to thermal fluctuations.
However, there is a well defined and pair-specific time $t_u$
at which the contact breaks for good, at least within
the unfolding time, $t_{\omega}$ that is defined by the
requirement that all non-local contacts are broken.
Specifically, the
non-locality refers to the sequencial distance $|j-i| > l$, where
$l=4$ (non-helical). 
The example values of $t_{\omega}$ are indicated in the
top panels of Figure 1.
The unfolding scenarios
may be defined in terms of a list of the times $t_u$ that are
averaged over several hundred different trajectories.\\

The values of $t_{\omega}$ vary across the trajectories and we
define $t_{\Omega}$ as their median (a simple average would
be ill-defined if there was no unfolding within a cutoff duration
in the time evolution). The lower left panel of Figure 1
shows the temperature dependence of $t_{\Omega}$ for three model
proteins including the I27 domain of titin. On lowering the
$\tilde{T}$, $t_{\Omega}$ grows rapidly,
faster than according to the Arrhenius law, suggesting
perhaps a Vogel-Fulcher-like divergence at a finite $\tilde{T}_0$.
However, a finite system such as a single protein
can give rise to a divergence only at $\tilde{T}$=0. This statement
also applies to folding times which generally have a U-shaped
temperature dependence with divergences at zero and infinity
with the former being Arrhenius-like \cite{Socci,biophysical}
(mean field theories may lead to different conclusions).
The $\tilde{T}$-dependence of $t_{\Omega}$ appears
to be consistent, to a leading order with the $\exp(A/T^2)$ law.
Interpretation of this law, and corrections to it, remain to be
elucidated. Our data do not rule out a power law divergence either.\\

We have found that varying the parameter $l$ between 4 and 10
affects $t_{\Omega}$ insignificantly. However reducing $l$ below 4
results in a substantially different $t_{\Omega}$ as shown in the
lower right panel of Figure 1 for the 1bba protein for $l=2$.
This suggests a physical relevance of $l=4$ for distinguishing
between local and non-local contacts \cite{Maritan}.
Considering $l$ smaller than 4 is impractical computationally
for proteins that are bigger than 1bba, and appears to have
no justification in the chemical denaturation \cite{Shortle}.
The lower right panel of Figure 1 shows that the divergence of the 
unfolding time also applies to secondary structures. In that
theoretical case, $l=2$ is a more sensible choice to take.\\ 

There are characteristic temperatures that are associated with the
processes of folding and stretching. In the case of folding, it is
the temperature of kinetic optimality (0.275 for 1tit). 
In the case of stretching, it is the temperature (0.8 for 1tit) at which
the purely entropic behavior \cite{entropic} sets in: above it, force
peaks disappear and the system acquires the worm-like-chain
behavior \cite{Edwards}. Is there a characteristic temperature,
$T_{\Omega}$ that can be associated with thermal unfolding?
Figure 2 suggests that there indeed such a temperature exists
and its reduced value for titin is around 1.1. Below this
temperature, the median
and the most probable unfolding times diverge
from each other significantly, indicating emergence of a broad 
distribution of time scales and temporal separation
of the unfolding events.\\

The left top panel of Figure 3 shows the average scenarios of the
unfolding events in titin at two temperatures:
at $T_{\Omega}$ and substantially below it, i.e.,
at $\tilde{T}$=0.85. These scenarios show the average unfolding
times, $t_U =<t_u>$ of specific contacts.
These times are plotted against the contact order, i.e., against
the sequential distance $|j-i|$. In order to see the details
in the scenarios, we actually plot $t_U - t_{\Omega}$ which removes
the dominant time scale. We observe that at and above $T_{\Omega}$,
the thermal fluctuations destroy bonds nearly simultaneously, independent
of the contact order. On the other hand, below $T_{\Omega}$,
the scenarios acquire reproducible structures in which the contacts
between strands C and F (solid squares) and between strands
A and G (solid circles) disintegrate much sooner, on an average, than
those between strands B-E (open triangles) and B-G (open circles).
This order of events agrees with all-atom nanosecond
long unfolding simulations
of titin by Paci and Karplus \cite{Paci} in which several trajectories
were studied at 450 K.\\

In broad features, the unfolding scenario at $\tilde{T}=0.85$ runs in
reverse to the folding scenario at $T_{min}$ shown in the top right panel
of Figure 3. However, the cross correlation plot between the two
scenarios, shown in the bottom left panel of Figure 3, indicates
that the two processes are not simply unticorrelated but merely
reflecting the time flow of events in a fairly monotonic fashion.
It is interesting to note that the events which are most relevant
to the search of the transition state --
the final stages of folding and the initial stages of unfolding --
anticorrelate in a nearly linear way. This point qualitatively agrees
with all-atom simulations for the $\beta$-hairpin fragment of 
protein G \cite{Pande2} that were performed between certain
characteristic sets of conformations (16 amino acids were considered,
the unfolding simulations took place at 350 K).\\

The folding scenario shown in the top right panel of Figure 3 is
defined in terms of average times, $t_c$, at which specific contacts
are established for the first time \cite{biophysical}. 
(Folding is considered to be fully accomplished when all contacts
are simultaneously established for the first time).
We have found 
that folding in the model titin takes place in two channels.
In the first channel, comprising about 24\%
of the trajectories, the C-F contacts are established in twice as long
a time as needed to set the A-G and A'-G contacts. In the second channel,
comprising the remaining 76\% of the trajectories
(at the temperature of optimal folding), C-F gets established
somewhat earlier than the A-G contacts. The scenario shown in Figure 11
of Reference \cite{thermtit} combines the two channels. The scenario
shown here discards the minority channel. Our studies of a generalized 
model of titin, in which the C$^{\beta}$ atoms are also included in the
description of the model, agrees qualitatively with the majority-channel scenario.
This updated scenario is consistent with the $\phi$-value 
data \cite{Fowler2001,Geierhaas}.\\

Finaly, we consider mechanical stretching of 1tit at constant speed.
Its scenario is defined in terms of the last average distance, $d_u$,
at which contacts are still holding when the C-terminus is moving
at a constant speed and the N-terminus is attached to an elastic anchor.
We have already
established that stretching at "room temperature" proceeds in a
way that is unrelated to folding \cite{Robbins,thermtit}. It is only
in the entropic limit, when stretching is governed exclusively
by the sequential distance that stretching is 
approximately reverse to folding at optimality \cite{entropic}.
The right bottom panel of Figure 4 shows that when unfolding and
stretching are both done at $\tilde{T}=0.85$ they follow each
other in a monotonic way.
The order of events is nearly identical
but the time intervals between them do not scale linearly except perhaps 
at the very begining of the two processes.
The reduced temperature of 0.85 belongs to the entropic regime
but, at the same time, it is below $T_{\Omega}$.
The inset of this panel shows, however, that stretching at
at the "room temperature" value of $\tilde{T}$=0.3 
does not correlate with unfolding at
$\tilde{T}$=0.85 at all.\\

In summary, we have provided an operational definition of the unfolding
times, demonstrated their "low"-temperature 
(faster than Arrhenius) divergence and indicated
existence of a characteristic temperature below which unfolding 
scenarios have contact-order-related structure 
and time scales become broadly distributed.
We have demonstrated that long time folding events are anticorrelated
with the short time unfolding events. 
We find that the simple topology-based
dynamical models qualitatively capture what is known from experiments and
simulations about the average order in which conformational changes
proceed in titin.\\

\section*{ACKNOWLEDGMENTS}
We appreciate discussions with Jane Clarke and Piotr Szymczak.
This work was funded
by the Ministry of Science in Poland (grant 2P03B 03225).


\newpage
\centerline{FIGURE CAPTIONS}

\begin{description}

\item[Fig. 1. ] 
The top panels show
examples of evolution of the distance, $d_{ij}$, between two contact
making amino acids $i$ and $j$ in the I27 (1tit) domain of titin 
when starting from the native structure
and then applying a temperature of $\tilde{T}$=1.1. The contacts
involved are between the $\beta$-strands C, F and B, G.
The A, A', B, C, D, E, F, and G strands in titin
correspond to the sequential segments
4-7, 11-15, 18-25, 32-36, 47-52, 55-61, 69-75, and 78-88 respectively.
This protein consists of altogether 89 amino acids.
The lower panels show $t_{\Omega}$ for the systems indicated.
The lower left (right) panel refers to calculations done with the
$l=4$ ($l=2$) criterion.
The data are generally based on at least 201 trajectories;
above $\tilde{T}$ of 0.8 -- on at least 501 trajectories.
There are two data sets for titin. The solid symbols correspond
to the Go-like model discussed in this paper whereas
the open symbols correspond to a generalized Go-like model
with side groups in which the degrees of freedom related to
the C$^{\beta}$ atoms are included. The generalized model shows
a similar behavior. 
The lines in the lower panel illustrate fits to the
$t_{\Omega}/\tau \; = \;\exp(A/{\tilde T}^2 -B/\tilde{T} + C)$ law,
where the sets ($A$, $B$,$C$)
are (10.592,10.492,6.976) for 1tit, (8.381,9.887,6.258) for 1crn,
(1.391,0.943,1.548) for 1bba with $l$=4, (5.259,1.576,4.916) for
1bba with $l$=2, (2.329,0.671,0.659) for the hairpin, and
(3.363,2.143,2.585) for the helix.
The fitting confidence level is at least 0.987.
Somewhat poorer fits were obtained by using the
$exp(exp(D/\tilde{T})$ law.
The data points for the $C^{\alpha}$-based model can also be 
superficially fitted to the Vogel-Fulcher-like divergences 
with the apparent $\tilde{T}_0$
of 0.56, 0.44, and 0.20 and with the energy
barriers of 2.6 2.3 and 1.5 for 1tit, 1crn, and 1bba respectively.\\

\item[Fig. 2. ] The distribution of unfolding times for 1tit
for the three temperatures indicated. The arrows point at the
median values. The inset in the middle panel shows the temperature
dependence of the difference between the median and peak
values in the distributions.\\

\item[Fig. 3. ] The top left and right panels show the unfolding and folding
scenarios in 1tit respectively. The data are averaged over 501 trajectories.
The unfolding scenarios are shown
for the two temperatures indicated. The broken line separates the
data points obtained for the two temperatures. 
The folding scenario corresponds to the temperature of the fastest
folding. The symbols assigned to specific contacts are the same in both panels.
Open circles, open triangles, open pentagons, solid circles, and
solid squares correspond to contacts B-G, B-E, D-E, A-G, and C-F
respectively. The stars denote all other contacts.
The lower left panel cross-plots the folding scenario with the
unfolding scenario at the temperatures indicated.
The lower right panel cross-plots events of mechanical stretching
with those of thermal unfolding.
The mechanical stretching in the inset and the
main panel is performed at $\tilde{T}=0.3$ and 0.85 respectively.\\

\end{description}

\begin{figure}
\epsfxsize=7in
\centerline{\epsffile{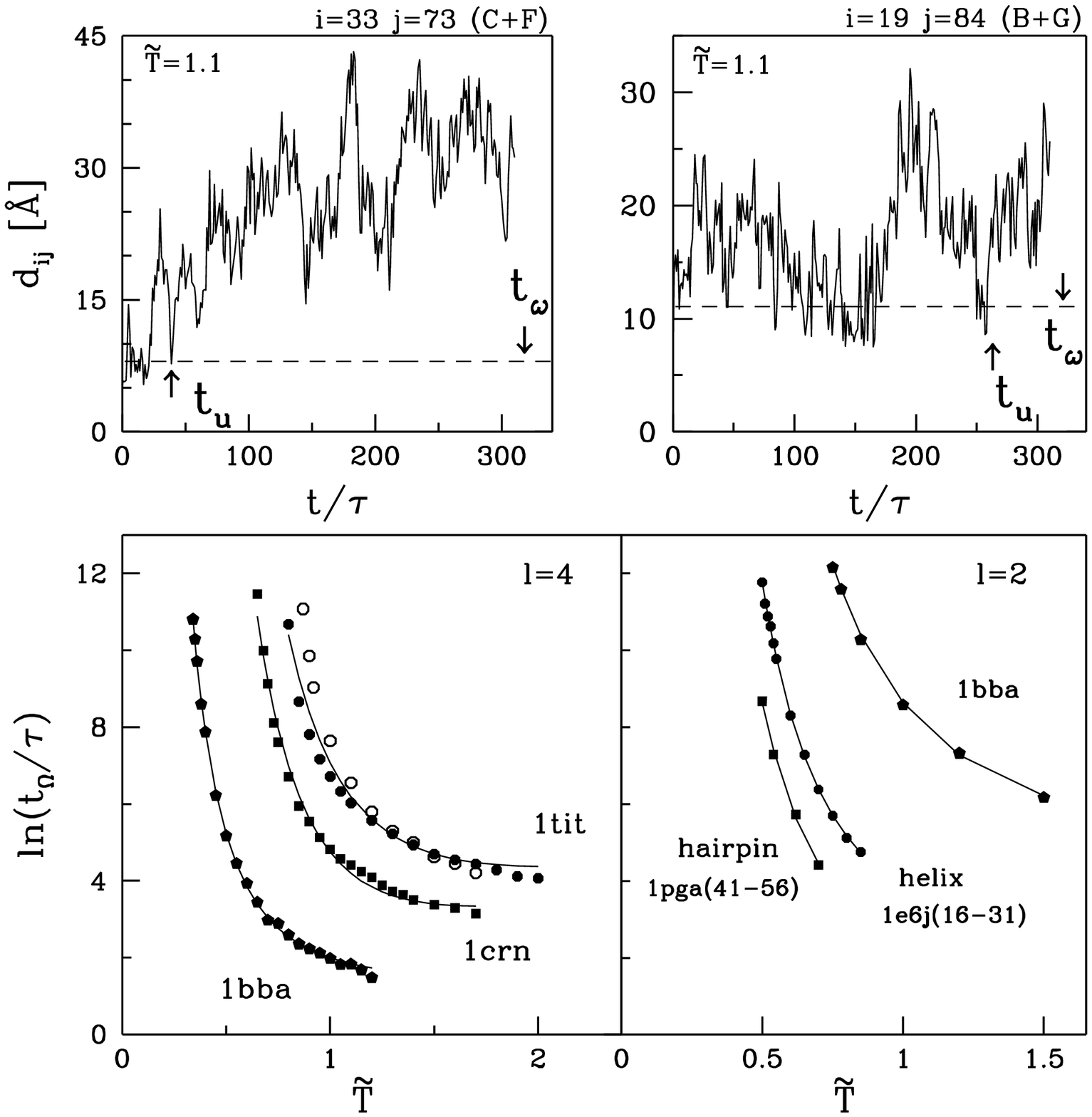}}
\vspace*{3cm}
\caption{ }
\end{figure}

\begin{figure}
\epsfxsize=7in
\centerline{\epsffile{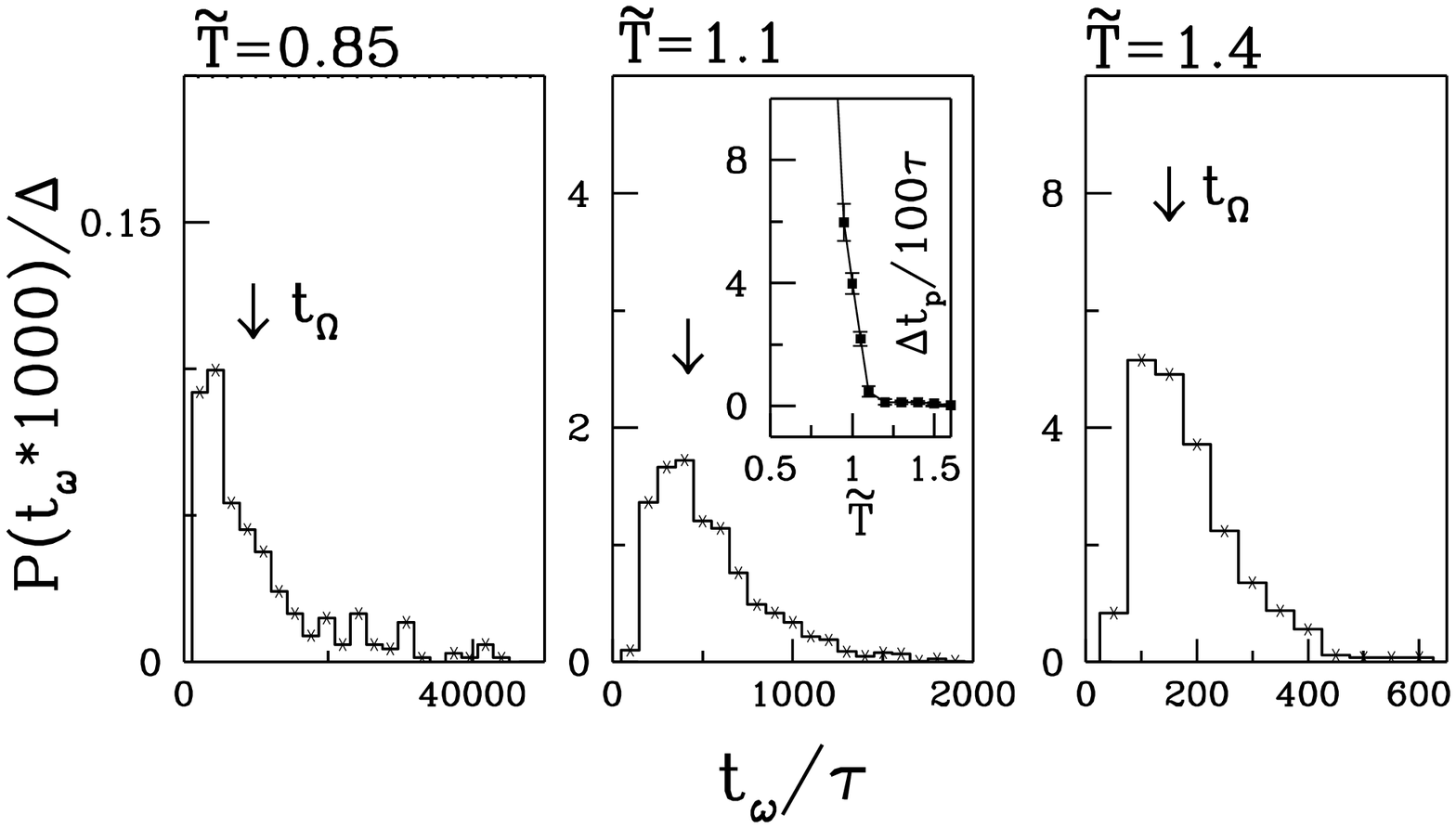}}
\vspace*{3cm}
\caption{ }
\end{figure}

\begin{figure}
\epsfxsize=7in
\centerline{\epsffile{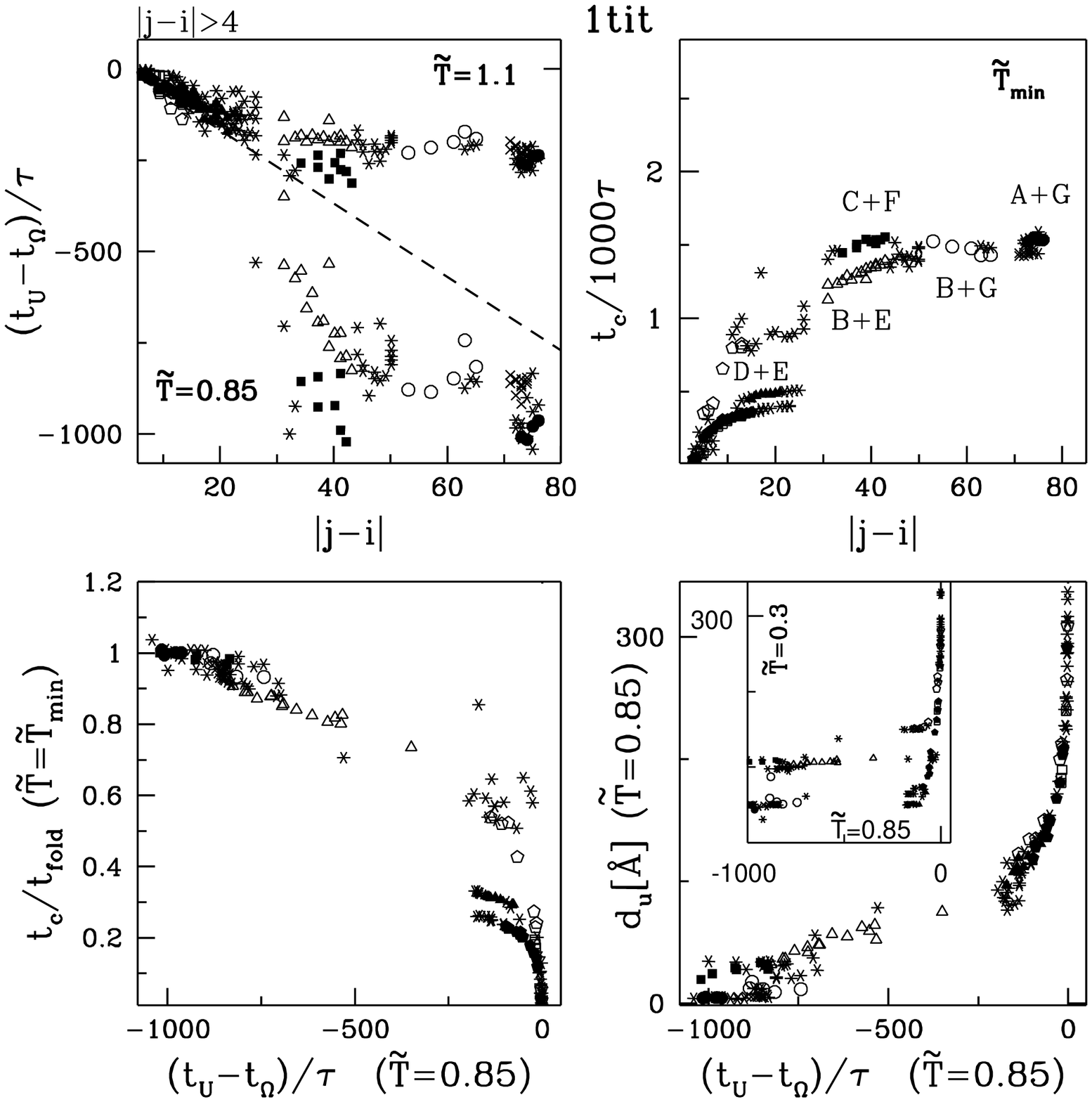}}
\vspace*{3cm}
\caption{ }
\end{figure}

\end{document}